\documentclass[aps,prb,reprint,showpacs]{revtex4-1}
\usepackage{amsmath,amssymb}
\usepackage{graphicx}
\usepackage{lmodern}
\usepackage{textcomp}
\usepackage[T1]{fontenc}

\begin{document}
\title{Apparent rippling with honeycomb symmetry and tunable periodicity observed by scanning tunneling microscopy on suspended graphene}
\author{A. Georgi$^1$, P. Nemes-Incze$^1$, B. Szafranek$^2$, D. Neumaier$^2$, V. Geringer$^1$, M. Liebmann$^1$, and M. Morgenstern$^1$}
\email[]{mmorgens@physik.rwth-aachen.de}
\affiliation{$^1$ II. Institute of Physics B and JARA-FIT, RWTH Aachen University, D-52074 Aachen, Germany\\ $^2$ Advanced Microelectronics Center Aachen (AMICA), AMO GmbH, Otto-Blumenthal-Strasse, D-52074 Aachen, Germany}

\date{\today}
\begin{abstract}
Suspended graphene is difficult to image by scanning probe microscopy due to the inherent van-der-Waals and dielectric forces exerted by the tip which are not counteracted by a substrate. Here, we report scanning tunneling microscopy data of suspended monolayer graphene in constant-current mode revealing a surprising honeycomb structure with amplitude of 50$-$200 pm and lattice constant of 10-40 nm. The apparent lattice constant is reduced by increasing the tunneling current $I$, but does not depend systematically on tunneling voltage $V$ or scan speed $v_{\rm scan}$. The honeycomb lattice of the rippling is aligned with the atomic structure observed on supported areas, while no atomic corrugation is found on suspended areas down to the resolution of about $3-4$ pm. We rule out that the honeycomb structure is induced by the feedback loop using a changing $v_{\rm scan}$, that it is a simple enlargement effect of the atomic lattice as well as models predicting frozen phonons or standing phonon waves induced by the tunneling current. Albeit we currently do not have a convincing explanation for the observed effect, we expect that our intriguing results will inspire further research related to suspended graphene.
\end{abstract}
\maketitle
\section{Introduction}
Two-dimensional (2D) materials such as graphene can be stabilized by an anharmonic coupling of in-plane and out-of plane phonons leading to a rippled 2D layer.\cite{Katsnelson}
Transmission electron microscopy (TEM) probing the suspended areas of partly supported graphene has found such a rippling on length scales of about 10 nm with amplitudes of about 1 nm by analyzing the tilt dependence of local diffraction patterns.\cite{Meyer} The length scales are in nice agreement with predictions from molecular dynamics (MD) simulations for freely suspended graphene.\cite{Vozmedanio}  Experimentally, rippling can also be induced by the substrate,\cite{Ishigami,Stolyarova, Cullen} by a non-conformal transfer of graphene onto a support material,\cite{Geringer,Mashoff,Morgenstern,Calado,Lanza} by heating and cooling \cite{Bao,Tapaszto,Bai} due to the negative thermal expansion coefficient of graphene,\cite{NTA}
by compressive edge stress,\cite{Shenoy} or by stress around defects.\cite{Warner} This has been recently reviewed in detail.\cite{Deng} \\
Probing the detailed topography of the rippling on suspended graphene is difficult. TEM suffers from a low contrast of monolayer graphene and its intrinsic low sensitivity to the vertical displacements,\cite{Meyer2,Wang,Warner,Bangert} while scanning probe microscopy modifies the structure due to the attractive van-der-Waals force exerted onto the graphene by the tip, which can be accompanied by attractive dielectric forces depending on the local tip potential.\cite{Mashoff,Stroscio,Xu1} This leads to a strong pulling of the graphene membrane by the tip, e.g., within constant-current mode imaging by a scanning tunneling microscope (STM). The pulling height was measured to be  about $50$ nm on suspended areas, if the tip is $1-7$~$\mu$m apart from the support.\cite{Stroscio,Xu1} Resulting lateral modifications of the graphene morphology by the tip forces have been observed rather directly within a dual tip STM.\cite{Eder}\\
Nevertheless, interesting effects have been found using STM on suspended graphene. For example, the tip forces induce an inhomogeneous pseudomagnetic field of about 10 T directly below the tip.\cite{Stroscio,Stroscio1} This leads to confinement of electronic states, which exhibit Coulomb blockade and orbital splittings in tunneling transport.\cite{Stroscio}
Atomic resolution and static ripples of 5-10 nm length scale have also been observed in suspended areas, if measured slowly and on small length scales.\cite{Zan,Stroscio,Xu1,Zhao}
In some experiments, it has been found that the amplitude of the atomic corrugation depends on scan speed and can be up to 1 nm, much larger than for supported samples.\cite{Xu1}
Moreover, the membrane retracts by up to 20 nm while increasing the current, which is claimed to be related to local heating,\cite{Xu2} albeit quantitatively not compatible with the large heat conductivity of graphene (see below).\cite{Pop}  Time series in constant current mode revealed a current dependence of the sub-Hz-frequency spectrum of the vertical tip movement \cite{Xu3} as well as occasional jumps up to 40 nm attributed to unbuckling processes of the membrane.\cite{Neek}
These results have recently been described by two power laws distinguishing vibrational motion and unbuckling processes.\cite{Peeters}\\
Here, we report a surprising effect that we observed occasionally on suspended graphene, namely an apparent rippling with honeycomb symmetry, aligned with the atomic lattice as observed on the supported areas, but with much larger period. Keeping the tunneling current constant, the rippling pattern is reproducibly observed on the identical area  independently on scan speed or scan direction of the tip. Furthermore, the period is increased from 10 nm to 40 nm  by decreasing the tunneling current from 500 pA to 100 pA.
The effect has not been reported in previous publications, maybe since these experiments are mostly performed on smaller membranes with diameters of 1 $\mu$m only and provide much smaller STM image sizes below $20\times 20$ nm$^2$.\cite{Stroscio,Stroscio1,Zan,Xu1,Xu2,Neek,Zhao}
Excluding several possible explanations, we could not come up with a convincing description of the apparent rippling, thus, hoping to stimulate further successful theoretical work.\\
\begin{figure}[bt]
\includegraphics[width=\linewidth]{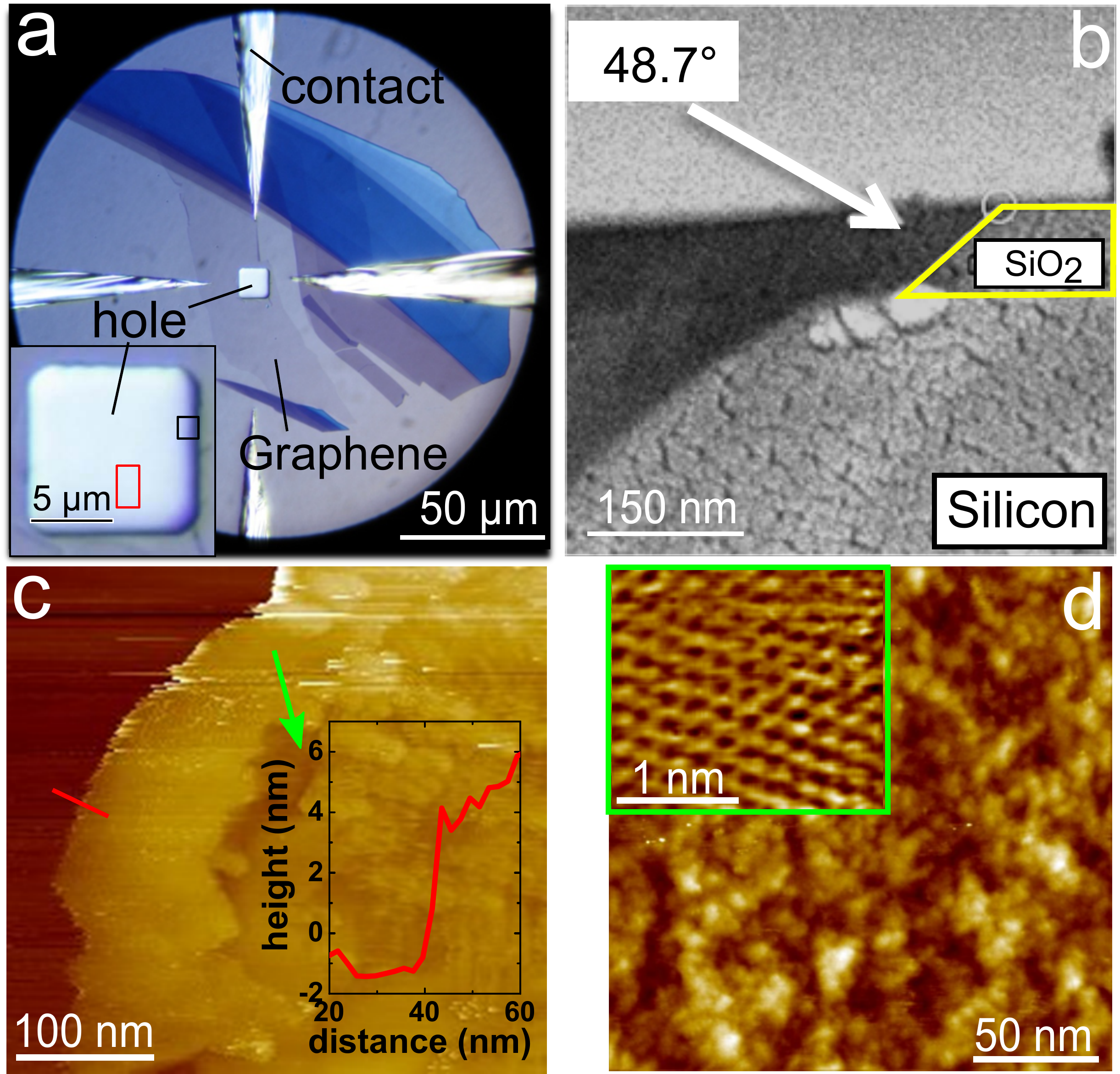}
\caption{(color online).
(a) Optical microscope image of graphene (light grey contrast) on  Si/SiO$_{2}$ partly suspended above a square marker hole ($10\times10\,\mu\text{m}^{2}$, light blue contrast); areas with stronger contrast are graphite; four indium leads act as contacts.\cite{Girit,Geringer2} Inset: Zoom into the area of the hole covered by graphene with encircled areas imaged in (c) (black square) and Fig. \ref{Fig2} (red rectangle). (b) Cross-sectional SEM image of a marker edge; 90\,nm thick SiO$_{2}$ is encircled in yellow exhibiting an angle of $48.7^\circ$ with respect to the surface. (c) STM constant-current image  recorded at the marker edge (black square in a), $I=100\,\text{pA}$, $V=-300\,\text{mV}$, scan direction from left to right; red line marks the position of the profile line shown as inset; green arrow marks the likely position of the real edge. (d) STM constant-current image of supported graphene area $(I = 200$\,pA, $V = -300$\,mV) with atomic resolution image as inset $(I = 300$\,pA, $V = -150$\,mV); c and d are measured consecutively on the different sample areas with the same tip.
\label{Fig1}}
\end{figure}
\section{Experiment}

Graphene is exfoliated on 90\,nm thick SiO$_{2}$ on Si(001) using adhesive tape.\cite{Novoselov} The substrate includes rectangular holes ($10\times10\,\mu$m$^2$) fabricated by reactive ion etching. By chance, a monolayer of graphene was deposited partly on such a hole (Fig.\,\ref{Fig1}a), i.e., it is suspended across $10\times10\,\mu$m$^2$. The hole edge exhibits a declining slope of $\approx 50^{\circ}$ with respect to the surface due to the etching process (Fig.\,\ref{Fig1}b). Atomic force microscopy (AFM) reveals additionally that some sections of the rim exhibit an outer area with a slope of only $14^{\circ}$. Previous AFM measurements found that the graphene sticks to the edges of such holes.\cite{Lee2008} Thereby it gains adhesion energy while being increasingly strained before eventually getting fully suspended within the center of large enough holes. The largest part of our graphene flake (Fig.\,\ref{Fig1}a) resides on the substrate fully encircling the suspended area. At its edges, it is surrounded by thicker graphite areas (dark blue areas in Fig.\,\ref{Fig1}a) .\\
Raman spectroscopy is used to verify that the suspended graphene is a monolayer.\cite{Ferrari} Afterwards, the graphene is electrically contacted by microsoldering with indium while at $170^{\circ}$\,C in ambient environment.\cite{Girit,Geringer2}
After transfering into a home-built ultrahigh vacuum STM operating at a temperature of $T=300$ K, the STM tip was aligned to the graphene flake using an optical microscope with a focal range up to 20 cm and a resolution of $\approx 10\,\mu$m \cite{Geringer2}. During this process, the clearly visible In leads act as additional cross hairs. The tip approach is then performed on the supported areas in order to avoid destruction of the suspended areas by tip forces. Afterwards we manoeuvered the tip to the suspended areas without loosing tunneling current. Later, we also approached on the suspended areas, but starting from smaller tip-sample distances.\\
STM images were measured in constant-current mode at scan speed $v_{\rm scan}$, current $I$, and sample voltage $V$. In order to avoid dielectric forces during imaging we operate at $V\simeq -200$ mV close to the contact potential. This compensates work function differences between tip and sample. The contact potential can be determined by relating $I(V)$ curves on suspended and supported areas as described elsewhere.\cite{Mashoff} $I(z)$ curves ($z$: vertical distance of tip) are recorded with feedback off after stabilizing the tip at current $I_{\rm stab}$ and voltage $V_{\rm stab}$.

\section{Results}
\begin{figure}[bt]
\includegraphics[width=6cm]{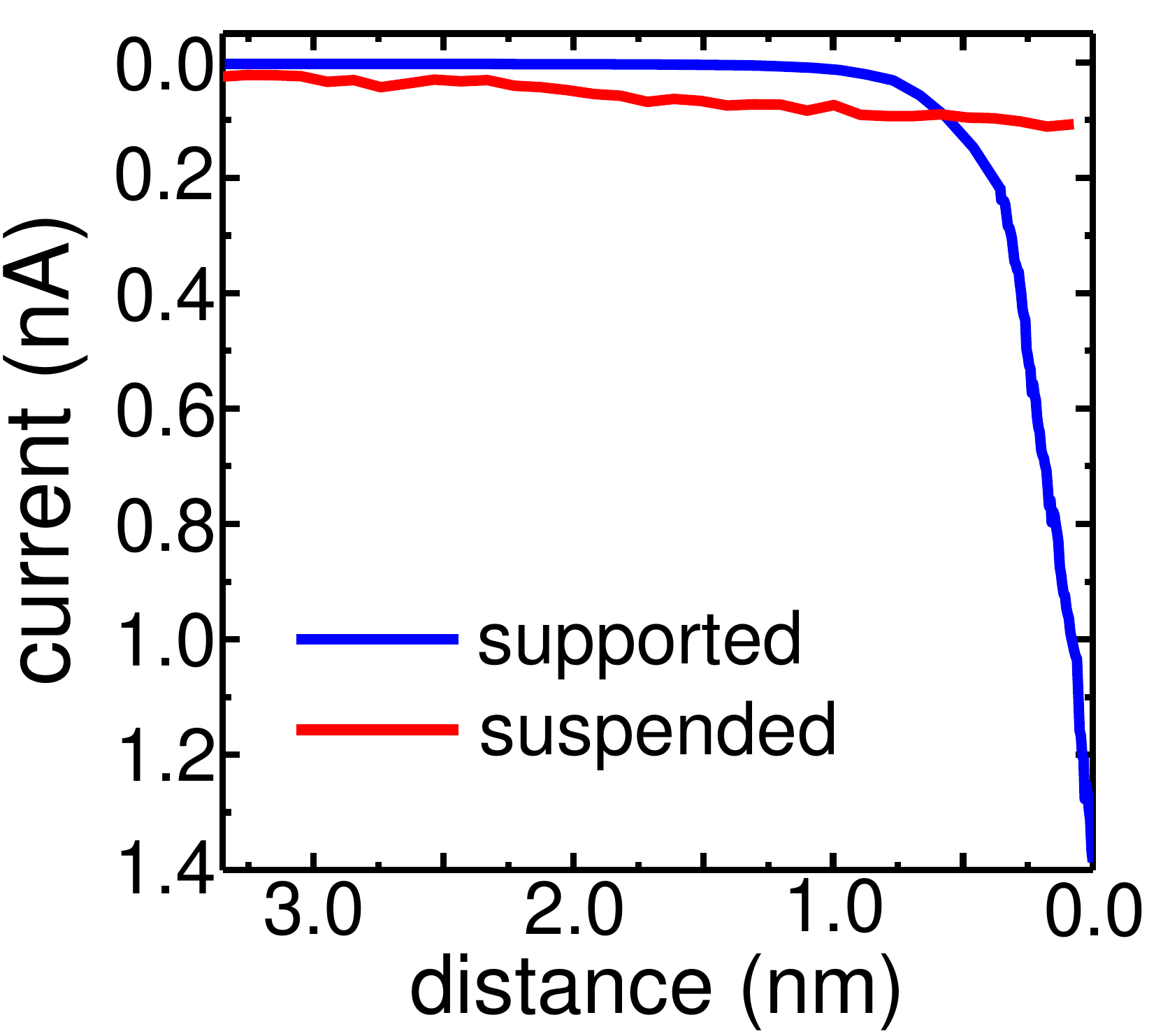}
\caption{(color online).
$I(z)$ curves  recorded on suspended graphene (area of red rectangle in inset of Fig. \ref{Fig1}a) and supported graphene as marked, $V_{\rm stab} = -300$\,mV, $I_{\rm stab} = 100$\,pA (suspended), $1.2$\,nA (supported).
\label{Fig1b}}
\end{figure}
\begin{figure*}[tb]
\includegraphics[width=13.0cm]{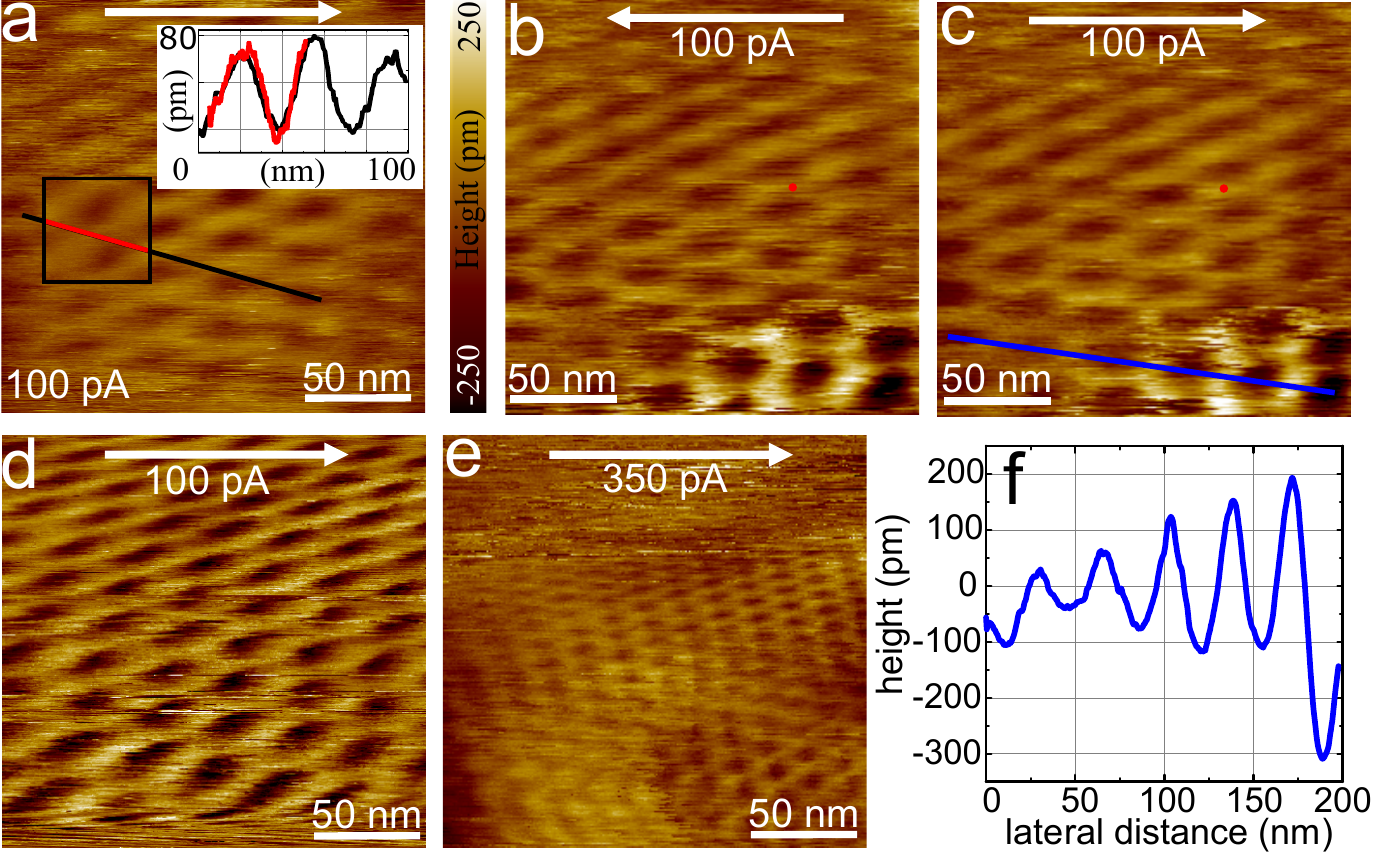}
\caption{(color online).
(a)-(e) STM constant-current images on suspended graphene recorded in part of the area marked by a red rectangle in the inset of Fig. \ref{Fig1}a; a linear upwards slope of $4^\circ$ in the direction from left to right is subtracted; the fast scan direction during recording is indicated by a white arrow marked by the tunneling current $I$ in b$-$e; same color scale in all images. (a) Main image: $I = 100$\,pA, $V = -300$\,mV,  $v_{\rm scan}$ = 600$\,$nm/s; small inset within black rectangle: subsequently measured image at $I = 100$\,pA$, $ $V = -300$\,mV,  $v_{\rm scan}$ = 20$\,$nm/s and higher resolution (factor 16) at the position, where it is overlayed; black and red lines mark the profile lines displayed in the inset. (b), (c) Subsequently measured images of the same area in scan directions as marked, $I = 100$\,pA,$ $ $V = -300$\,mV,  $v_{\rm scan}$ = 600$\,$nm/s, blue line marks profile line displayed in f; the two images are recorded simultaneously during forward scan and backward scan, i.e the same vertical image position is used for both horizontal scans before the tip is moved to the next vertical position; red dots mark identical positions within the image. (d) Image of the same area measured after vertically retracting the STM tip from the graphene membrane by about 100 nm and reapproaching it; same parameters as in b, c. (e) Same area measured with the same parameters as in b-d except $I=350\,$pA. (f) Profile line along the blue line in c.
\label{Fig2}}
\end{figure*}
Figure \ref{Fig1}a shows an optical image of the graphene flake completely covering the marker hole.
The colored frames in the inset mark the areas where STM measurements have been pursued. The slope of the marker hole is mostly about 50$^\circ$ (Fig. \ref{Fig1}b), while the slope of the graphene as measured by STM on the suspended regions is typically 4$^\circ$ revealing that the graphene is indeed suspended above the marker hole. Figure \ref{Fig1}c shows an STM image obtained close to the edge of the marker hole while scanning from left to right. It exhibits a sharp step of 4 nm height.
On the right side of the step, the graphene appears relatively flat (rms roughness: $\sigma= 0.3$ nm on 50 nm length scales,  $\sigma= 1.3$ nm on larger length scales) until we observe an additional hole-like area (green arrow) followed by an area of stronger corrugation also on smaller length scales (rms roughness: $\sigma = 1.7$ nm). The later value of $\sigma$ is much larger than the roughness observed far away from the marker hole edge. The latter is $\sigma=0.25$ nm (Fig. \ref{Fig1}d) being in good agreement with earlier results of  graphene on SiO$_2$.\cite{Geringer,Cullen, Stolyarova,Ishigami} The large roughness is most likely induced by a rougher substrate around the marker hole caused by the etching process. The sharp step is not observed while scanning from right to left, while the area to the right of the step is imaged identically (not shown). We assume that the step marks an abrupt delaminating of the graphene from the edges of the hole due to the attractive van-der-Waals forces of the tip, while the increased corrugation marks the onset of supported graphene.
Most importantly, the transition from supported to suspended graphene as marked by the green arrow can be determined with a precision of, at least, 100 nm.\\
A much weaker decay of $I(z)$ curves is observed on the suspended areas (Fig. \ref{Fig1b}). The current decays nearly linearly by about 25 pA/nm starting from 100 pA at closest distance. This indicates that the membrane is pulled with the tip (see below). We find featureless $dI/dV$ curves on both areas with minima around $V=0$ mV (not shown).\\
Figure \ref{Fig1}d shows STM images on the supported areas. They are recorded more than  1 $\mu$m away from the edge of the marker hole. They exhibit the well known rippling of graphene down to 10 nm length scales with an rms roughness of 0.25 nm.\cite{Geringer,Cullen, Stolyarova,Ishigami} Atomic resolution with the typical honeycomb lattice and a corrugation of about $30$ pm is observed on smaller length scales (inset).\\
\begin{figure*}[bt]
\includegraphics[width=12.0cm]{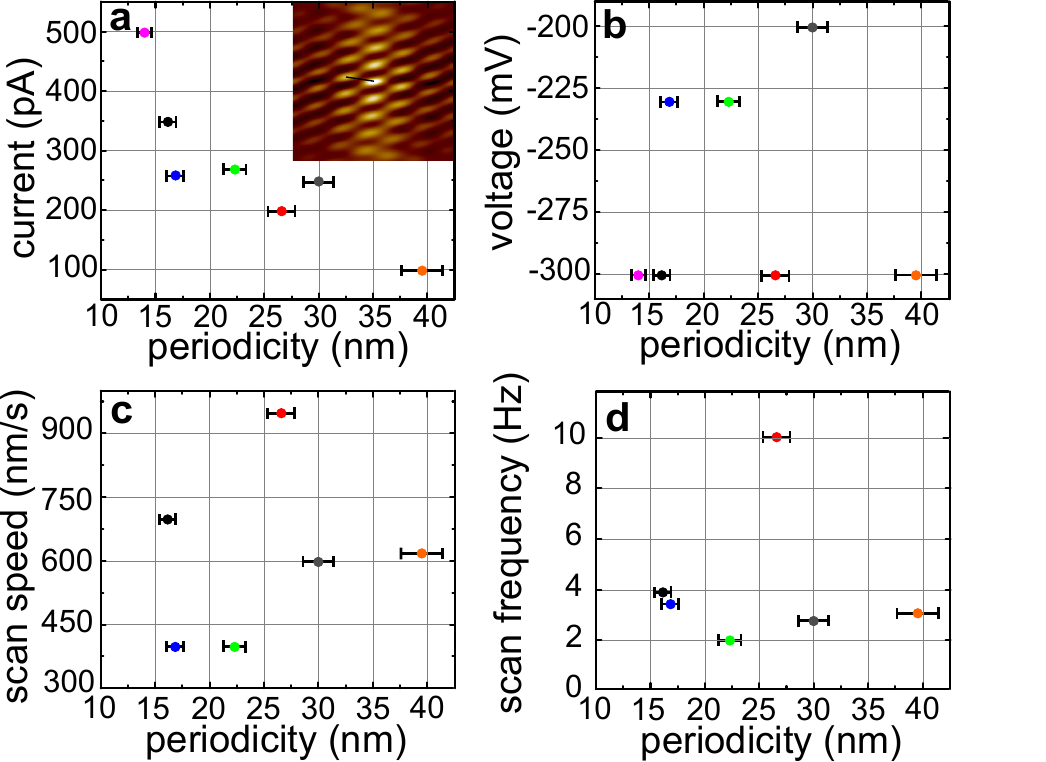}
\caption{(color online).
(a) Periodicity of the corrugation of the superstructure along the fast scan direction as a function of tunneling current $I$. Inset: 2D autocorrelation of the STM image displayed in Fig. \ref{Fig2}a; such autocorrelation images are used to determine the periodicity, which is marked as a black line. Error bars are deduced from the fluctuation of periodicity along different scan lines within an image. (b) Periodicity of the same images as analyzed in (a), displayed in the same color, but as function of sample voltage $V$; note that $V$ is only moderately changed in order to avoid influences of dielectric forces. (c) Periodicity of same images as in a, b, but displayed as a function of scan speed $v_{\rm scan}$; (d) Periodicity of same images as in a, b,c, but displayed as a function of line frequency during scanning.
\label{Fig3}}
\end{figure*}
Figure \ref{Fig2} shows STM images obtained on the suspended areas more than 1 $\mu$m away from the edges of the marker hole. STM images recorded closer to the edge exhibit rather irregular structures and are quite unstable, i.e. the image appearance changes rapidly during scanning. But far away from the edges, we reproducibly observe a regular honeycomb structure.
Figure \ref{Fig2}a-c show subsequently obtained images on the same area revealing the reproducibility of the pattern. The inset enframed in black in Fig. \ref{Fig2}a is a zoom into that area recorded at much smaller $v_{\rm scan}$ revealing that the pattern is rather exactly reproduced independent of $v_{\rm scan}$, if tip, $V$ and $I$ are not changed (see also profile lines in the upper inset). Also the direction of scanning does not change the pattern except minor details (Fig. \ref{Fig2}b and c). We find a small relative offset of the patterns in the fast scan direction of about 10 nm, but no offset in the slow scan direction. An offset of $~5$ nm in the fast scan direction is also observed on supported areas using the same image size and the same $v_{\rm scan}=600$ nm/s and is traced back to the piezo creep of the tube scanner. Thus, most of the offset between Fig. \ref{Fig2}b and c is due to creep, but there might be an additional offset due to lateral drag of the graphene membrane by the tip of about 5 nm.\\
Finally, retracting the tip from the membrane in $z$-direction by more than 100 nm, thereby, completely loosing the tunneling current, and reapproaching, does barely changes the observed pattern. Figure \ref{Fig2}d shows an image of the same area as recorded in Fig. \ref{Fig2}a$-$c after such a new approach.\\
The fact that the very same tip revealed reasonable corrugation and atomic resolution images on supported areas (Fig. \ref{Fig1}d) prior, in between, and after the imaging of the large scale corrugation on the suspended areas excludes that the observed pattern is an artifact of the measurement equipment. We also carefully checked the error signal $\Delta I/I$ of the feedback loop being typically below 2 \% during imaging of the structures in Fig. \ref{Fig2}. Moreover, the tube scanner did not operate in its extreme positions during the imaging. Thus, we conclude that the observed patterns represent the real $z$-movement of the tunneling tip in constant-current mode, exhibiting a large scale corrugation with honeycomb symmetry.\\
The amplitude of the corrugation is changing within an image and also in between images by up to a factor of three (see e.g. line scan in Fig. \ref{Fig2}f) without any systematic dependence on measurement parameters. We believe that it is related to details of the tip shape, which are known to rapidly change at room temperature.\\

Most surprisingly, the periodicity of the apparent honeycomb lattice changes dramatically, if the tunneling current is changed. Figure \ref{Fig2}e is measured on the identical area as Fig. \ref{Fig2}a$-$d, but at a current of $I=350$ pA instead of $I=100$ pA, while maintaining $V$ and $v_{\rm scan}$. The period of the honeycomb lattice decreases from 40 nm to about 15 nm, while the amplitude barely changes being on average 100 pm at $I=350$ pA and 150 pm at $I=100$ pA.\\
Figure \ref{Fig3} shows this trend quantitatively for images recorded at slightly different positions within the red rectangle of the inset of Fig. \ref{Fig1}a. For these images, also $V$ and $v_{\rm scan}$ are slightly changed. The periodicity is deduced using autocorrelation maps of the constant current images as shown in the inset of Fig. \ref{Fig3}a. It is chosen as the distance between the central peak and the neighboring peaks closest to the direction of fast scanning (black line in inset of Fig. \ref{Fig3}a). These peaks are selected, since they are the least influenced by thermal drift or piezocreep. A clear trend of increasing period of the honeycomb lattice with decreasing current is found (Fig. \ref{Fig3}a), but no systematic dependence on voltage, scan speed, or scan frequency, i.e. the number of lines recorded per second (Fig. \ref{Fig3}b$-$d, see also Fig. \ref{Fig2}a for a more extreme case of changing $v_{\rm scan}$). The amplitude of the honeycomb corrugation varies between 50 pm and 200 pm, but again without a trend with respect to $I$, $V$, and $v_{\rm scan}$.\\

\begin{figure}[bt]
\includegraphics[width=8.5cm]{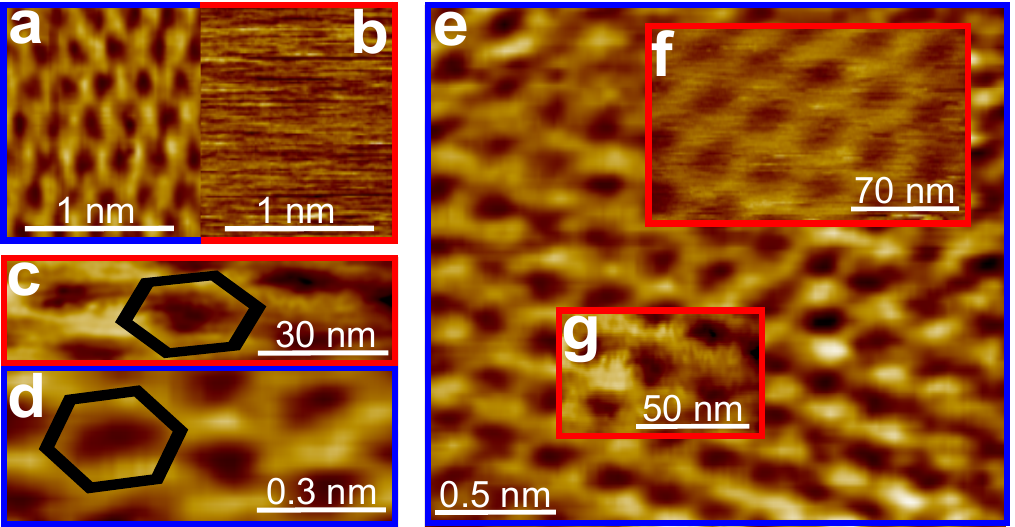}
\caption{(color online).
(a) Constant-current image of supported graphene exhibiting atomic resolution, $I=300\,\text{pA}$, $V=-150\,\text{mV}$. (b) Small scale constant-current image of suspended graphene area which exhibits large-scale corrugation,  $I=100\,\text{pA}$, $V=-300\,\text{mV}$, same lateral scale, and same height contrast as in a. (c) Constant-current image of suspended graphene, $I=100\,\text{pA}$, $V=-300\,\text{mV}$, honeycomb is overlaid. (d), (e) Constant-current images of supported graphene, $I=300\,\text{pA}$, $V=-150\,\text{mV}$, honeycomb is overlaid in (d). (f), (g) Constant-current images of suspended graphene, $I=100\,\text{pA}$, $V=-300\,\text{mV}$, stretched in vertical direction by a factor of 1.63 (f) and 1.49 (g) in order to compensate for presumable drift effects in that direction. The height scale (color scale) in c$-$g is adapted to obtain a similar contrast, i.e. the atomic resolution images are displayed with a threefold magnified contrast.
\label{Fig4}}
\end{figure}

Interestingly, the orientation of the long-range corrugation is quite similar to the orientation of the atomic lattice observed on supported areas with the same tip. In addition, the suspended areas do not exhibit atomic resolution. These results are emphasized in Fig. \ref{Fig4}. Figure \ref{Fig4}a and b compare STM images on supported and suspended areas using the same lateral magnification and the same color scale for the height information. While an obvious honeycomb lattice with the period expected from the atomic lattice of graphene is visible on the supported areas (corrugation amplitude: 30 pm), there is no honeycomb pattern apparent on the suspended areas down to the noise level of 5 pm. Even after removing the noise by adequate filtering we do not observe any signs of an atomic scale honeycomb pattern, such that we can exclude it down to amplitudes of 3$-$4 pm. Figure \ref{Fig4}c and d compare the large scale corrugation on suspended areas (c) with the atomic resolution on supported areas (d) by adjusting the magnification and the height contrast. The patterns are quite similar albeit the long range corrugation appears more strongly distorted, i.e. the lattice constant in horizontal direction is significantly larger than in vertical direction. If we symmetrize the honeycomb structure on suspended areas by stretching the images vertically, in order to compensate for likely effects of thermal drift, remaining piezocreep or drag, and overlap the resulting images (Fig. \ref{Fig4}f and g) on an atomic resolution image (Fig. \ref{Fig4}e), the similarity becomes even more striking.
\section{Discussion}
Albeit, we do not have an explanation for the observed effect, we will present the arguments against the most obvious possible scenarios.
\subsection{Instrumental artifacts}
First, we should exclude that the observed corrugation is an artifact of the measurement equipment, i.e.,  we simply measure the atomic corrugation on the wrong lateral scale. Such a scenario appears appealing since other authors find atomic resolution on suspended graphene with relatively large amplitudes of 0.1$-$1 nm.\cite{Zan,Stroscio,Xu1,Zhao} These amplitudes are similar to the amplitudes of our large scale corrugation. However, firstly we observed the large scale corrugation with different microtips (up to 3 month in between) on different areas  of the suspended region, e.g., within the red rectangle but also at the left edge of the black rectangle of the inset of Fig. \ref{Fig1}a. Moreover, we find it reproducibly prior and after moving to the substrate area, where we obtain normal STM images of graphene with atomic resolution as in Fig. \ref{Fig1}d. We also took care that the tube scanner has operated in its intermediate low voltage position with respect to $x,y$ movement and $z$ movement. Coming from the substrate, we had to follow a 4$^\circ$ downwards slope towards the center of the membrane indicating a remaining attraction of the graphene by the substrate most likely of van-der Waals type. Thus, the tube scanner has been extended by about 200 nm for the measurements on the membrane, still being well apart from the maximum extraction of 600 nm. Indeed, the tip was only extracted by $\pm 50$ nm with respect to its equilibrium position during all measurements on the membrane.
Thus, we can safely exclude a wrong operation of our instrument exclusively on the suspended areas.
\subsection{Model of the membrane during tunneling}
\begin{figure}[bt]
\includegraphics[width=8cm]{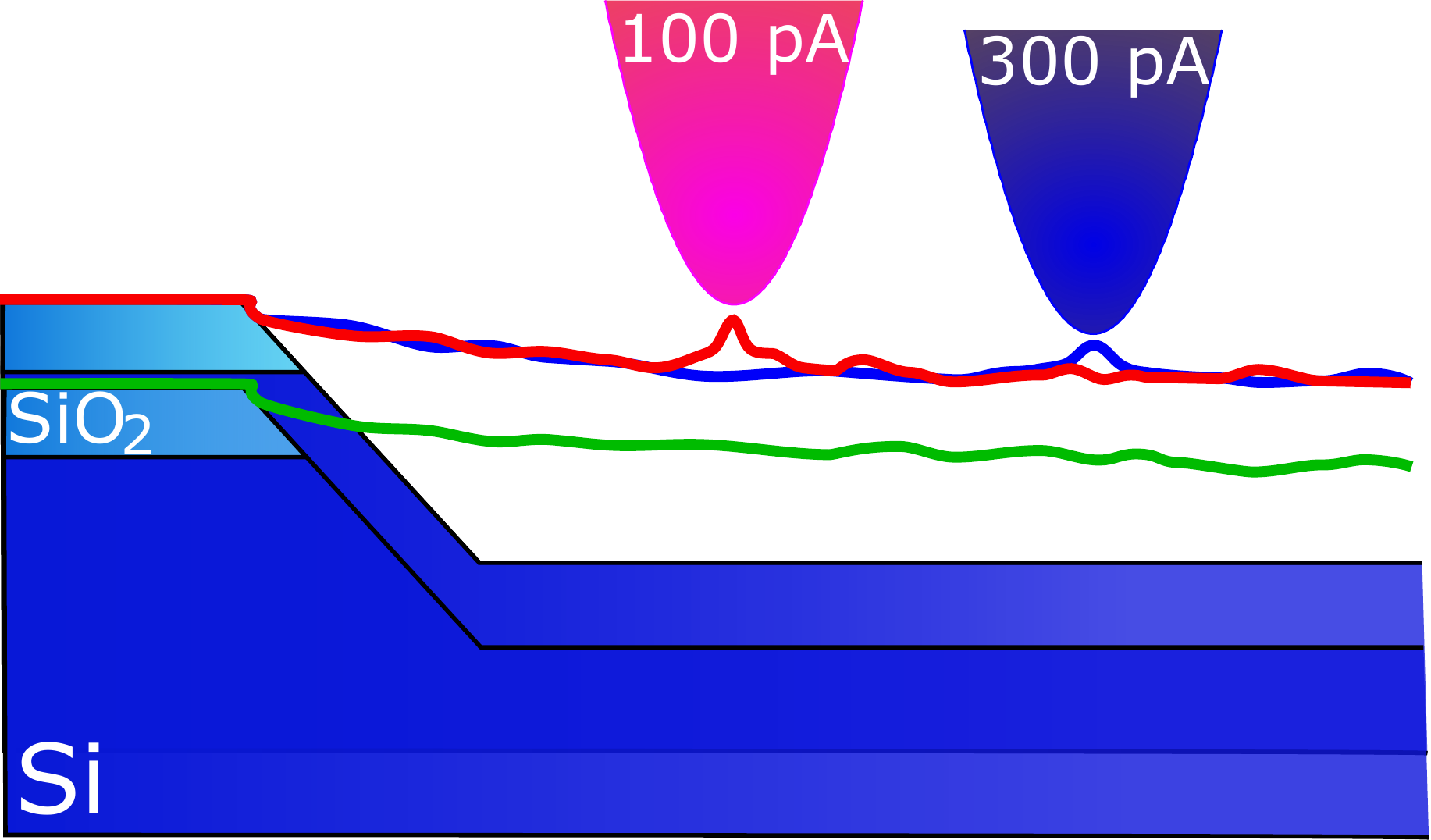}
\caption{(color online).
Sketch of suspended graphene without tunneling tip (green) and with tunneling tip at different currents as marked (blue, red) above the marker hole patterned into Si/SiO$_2$. Tunneling distances as well as amplitude and wavelength of the dynamic rippling of graphene are overemphasized for the sake of visibility.
\label{Fig6}}
\end{figure}

In order to discuss further possible explanations of the found apparent rippling, we need a description of the membrane during tunneling. Our model is sketched in Fig. \ref{Fig6}. It is largely based on previous, partly coarse grained MD  simulations of a freely suspended membrane in presence of a tip.\cite{Mashoff,Stroscio,Stroscio1,Nemes} These calculations revealed, that an Ir sphere (representing the tip) on top of a graphene membrane without a substrate below, but with a counteracting gate force, leads to a Gaussian bump within the graphene directly below the tip. The bump is surrounded by a concave shape of the membrane.\cite{Stroscio,Stroscio1} Below, we argue that the intrinsic rippling of the membrane is barely changed by the tip forces. We sketch this situation by the red and blue line in Fig. \ref{Fig6}. According to the experiments, we have to assume that the stretched area gets pulled with the tip, while lowering $I$, which is displayed by the difference between red and blue line.\\
We now describe the arguments leading to Fig. \ref{Fig6} in detail.\\

\begin{figure}[bt]
\includegraphics[width=8cm]{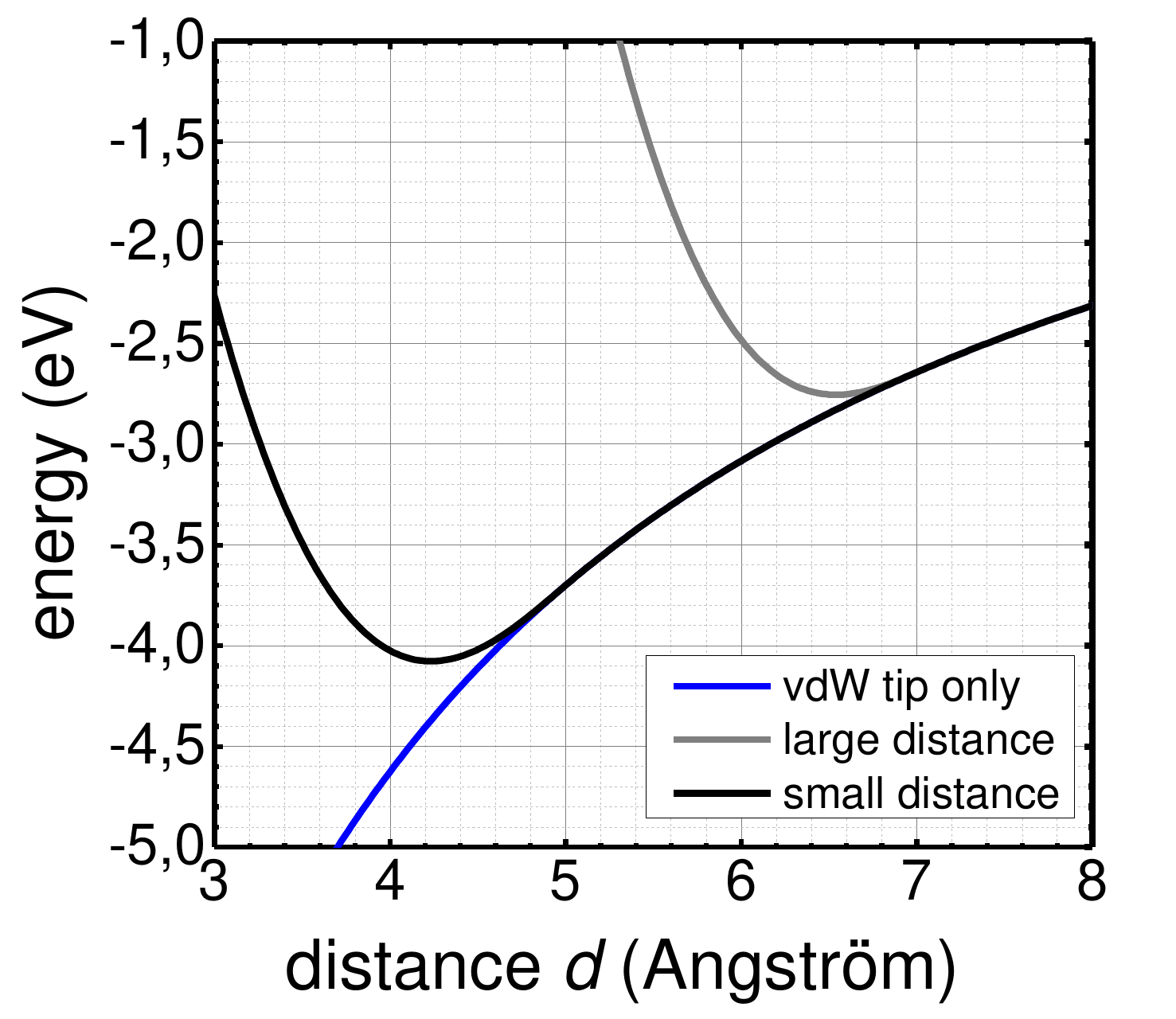}
\caption{(color online).
Estimated energy of the membrane as a function of shortest distance $d$ relative to the fixed tip. Blue line shows the attractive van-der-Waals energy provided by the tip $E_{\rm vdW,tip}(d)$. Grey and black lines show a superposition of $E_{\rm vdW,tip}(d)$ and the strain energy $E_{\rm strain}(h)$ within the membrane for $d(h=0)=7$ \AA\hspace{1mm} (grey) and  $d(h=0)=5$ \AA\hspace{1mm} (black) using $h=d(h=0)-d$ (see text).
\label{Fig7}}
\end{figure}

Firstly, the fact, that the tip used on the membrane reveals an exponential increase of the tunneling current up to $I>1$ nA on supported areas (Fig. \ref{Fig1b}), implies that we operate in tunneling distance at $I=0.1-0.5$ nA on suspended areas, too.
It excludes, in particular, the presence of an oxide barrier at the front of the tip, which would lead to tunneling even if tip and graphene are in contact, but would result in a saturation of $I(z)$.
It is very unlikely that the resistance of the graphene membrane itself increases up to 1-2 G$\Omega$ at 300 K due to stretching by the tip, while the tip is in ohmic contact with the membrane. In particular, such a high membrane resistance would lead to instabilities of the transport due to local charging by the tip electrons,\cite{Bindel} which was not observed in the experiment in agreement with earlier results.\cite{Stroscio}\\
In favor of the tunneling model without contact, one can easily show that the graphene membrane can reside in an energy minimum at tunneling distance of $d=4-6$ \AA. One firstly estimates the van-der-Waals energy $E_{\rm vdW,tip}$ between tip and graphene by the Hamaker approximation. We use a sphere with radius $R$ as tip in distance $d$ above a plane representing the graphene,\cite{Israel}
which reveals for $R>d$:
\begin{equation}
E_{\rm vdW,tip}= \frac{\pi C_{6A} \rho_{\rm W} \rho_{\rm G} R}{6d}
\label{eq:vdW}
\end{equation}
The required parameters to calculate the coefficient $C_{6A}$ and the atomic densities $\rho_{\rm W}$ and $\rho_{\rm G}$ of the W tip and graphene, respectively, are taken from the literature.\cite{parameters}
A sphere of $R=2$ nm matches the strength of the van-der-Waals energies revealed from MD simulations within the LAMMPS code,\cite{LAMMPS} thereby, using a W pyramid with (110) axis as tip. With this sphere, we obtain $E_{\rm vdW,tip}=-18.5$~eV\AA/$d$ [\AA].\\
There are a number of possible errors in this estimate. Firstly, the experimentally based Hamaker constants between graphene and W vary by up to a factor of two.\cite{parameters} Secondly, the tip radius could be different. It is bound by the fact that we observe atomic resolution, i.e., $R\le2$ nm, but we cannot exclude that the tip is effectively sharper   than the W pyramid leading to an effective radius down to $R\simeq 0.8$ nm. This reveals another factor of 2.5 of possibly larger van-der-Waals interactions. Thirdly, the model neglects details of the change of the polarizibility functions by the atomic environment.\cite{Tkatchenko1} The adapted $C_{6A}$ values result on the one hand from polarizability measurements of a W tip \cite{parameters} which should mimic the experimental situation well. On the other hand, the used values for C origin from Hartree-Fock calculations of Graphite adsorption energies \cite{parameters} and, thus, might also be close to the experimental ones. Corresponding errors are probably in the range of several 10 \%. Fourthly, screening is neglected, which reduces the van-der-Waals forces, again in the range of a few 10 \%.\cite{Tkatchenko2} Fifthly, Casimir-Polder forces caused by retardation are neglected, but they start to be of relevance at distances of more than 10 nm between the interacting materials only.\cite{Israel,Casimir} Finally and most importantly, long-range excitations can lead to an even qualitatively different behavior. The interaction can be significantly enhanced and has been found theoretically to vary by up to a factor of seven between different C structures notably exhibiting the largest effective, atomic $C_{6A}$ value for graphene. \cite{Tkatchenko3} An increase of atomic $C_{6A}$ values for C based molecules with molecular size by up to 20 \% is found experimentally, which confirms such a scenario.\cite{Tautz} Density functional theory based calculations even predict that decay exponents are changed due to the long range excitations, e.g. two graphene layers at about 1 nm distance exhibit an exponent of $-3.5$ instead of $-4$.\cite{Ambrosetti}
Unfortunately, accurate AFM measurements of force-distance curves absorb varying exponents in the unknown shape of the tip apex.\cite{AFM}
Thus, without detailed calculations, the resulting error can barely be estimated. However, due to the fact that our parameters are based on experimental values of a W tip and Hartree-Fock calculations of graphite make us confident that the corresponding error is not too large. So, we expect errors in the van-der-Waals forces between tip and graphene by up to a factor of five, which is the largest error in all the estimated forces in this manuscript.\\
The elastic strain energy within the graphene membrane due to stretching by the tip forces is also estimated by LAMMPS calculations. Therefore, we simulate graphene on
SiO$_2$ in presence of the pyramidal W tip. We use the adaptive imtermolecular reactive bond order (AIREBO) potential for the interatomic forces in graphene.\cite{AIREBO} It is crosschecked by continuum plate mechanics calculations using the known Lamé parameters of graphene.\cite{Lame}
The calculations confirm that the pyramidal W-tip induces a Gaussian bump within the graphene directly below the tip (even on the substrate).
This bump  exhibits a $\sigma$-width of 5-7~\AA.\cite{Nemes}
A similar width of the Gaussian within the concave areas has been found by MD simulations of a free membrane in presence of a tip.\cite{Stroscio,Stroscio1}
The corresponding strain energy $E_{\rm strain}$ as a function of the height $h$ of the Gaussian is well fitted by $E_{\rm strain}[eV]\simeq 0.6\cdot h$[\AA]$^{2.7}$ for typical $h=1-2$~\AA.\\
It is now straightforward to determine the potential energy for different $d$ as shown in Fig. \ref{Fig7}. We find pronounced metastable minima, e.g., at $d=4.2$ \AA\hspace{1mm} or $d=6.5$ \AA. The energy barrier towards the global minimum, where the graphene is attached to the tip, amounts to 10 eV and 60 eV, respectively. Consequently, the tunneling position is quite stable even at room temperature and even considering the large error bar of the tip-graphene van-der Waals forces.\\
 We conclude that the huge elastic modulus of graphene ($E=340$ N/m)\cite{Lee2008} allows stable tunneling on suspended graphene, which in turn requires that the membrane below the tip is stretched during tunneling. Corresponding maximum strains amount to about 1 \% according to the MD simulations.\cite{Nemes,Stroscio, Stroscio1}\\
 Notice that the equilibrium graphene-tip distance $d$ in Fig. \ref{Fig7} changes by $\delta d = 2.3$ \AA\hspace{1mm}, when $d(h=0)$ changes by only 2 \AA. This implies for the experiment, that the membrane  moves towards the tip by 0.3 \AA, when the tip is approaching by 2 \AA, such that the $I(z)$ curve on the membrane must be steeper than on supported areas.\cite{Mashoff} This is in obvious contrast to the experiment in Fig. \ref{Fig1b} and will be discussed below.

Secondly, we discuss the van-der-Waals forces of the substrate. The Si substrate is at a distance of $D=1.6$ $\mu$m below the membrane as determined by AFM. The energy density of its interaction with the graphene can be estimated\cite{Israel} using $\epsilon_{\rm vdW,G-Si}=\pi C_{6B} \rho_{\rm Si} \rho_{\rm G}/D$ with $C_{6B}$ taken from the literature,\cite{parameters} revealing  $\epsilon_{\rm vdW,G-Si}=13$ meV/$\mu$m$^2$. Thus, $\epsilon_{\rm vdW,G-Si}$ is irrelevant with respect to $E_{\rm vdW,tip}$ and $E_{\rm strain}$. The resulting total van-der-Waals energy across the membrane of  $E_{\rm vdW,G-Si}=1.2$ eV leads to minimal stretching of flat graphene. The strain amounts to less than 1 \AA\hspace{1mm} across the 10 $\mu$m of the membrane. However, in case that the membrane is larger than the hole, e.g., due to the heat treatment during sample preparation,\cite{NTA} the membrane could be bent downwards to the substrate due to the size mismatch and $\epsilon_{\rm vdW,G-Si}$. We believe that this explains the about $4^\circ$ downwards movement of the tip towards the center of the membrane.\\
Notice that the thermal energy of the out-of plane movement of the graphene, being close to $k_{\rm B}T =25$ meV/atom ($k_{\rm B}$: Boltzmann constant) at room temperature,\cite{ZPA} is much larger than $\epsilon_{\rm vdW,G-Si}$. Consequently, the graphene will be additionally dynamically rippled.\cite{Vozmedanio}\\
The tip will easily pull the graphene upwards against $\epsilon_{\rm vdW,G-Si}$. However, since the tip has typically a conical shape on larger scales with angles of about $20-30^\circ$ with respect to the axis of the tip wire, there will be large areas of the suspended graphene where the substrate is closer to the graphene than any tip area (see Fig. \ref{Fig6}). Thus, we expect that the graphene still globally hangs downwards, while only pulled upwards within an area of about $1\times 1$ $\mu$m$^2$ below the tip. This explains that we have to move downwards by about $4^\circ$ towards the center of the membrane.
\\
Of course, there are also errors in the determination of the van-der-Waals forces between graphene and the substrate. However, since the parameters are matched to experiments, they are probably not too large, such that the general conclusions are not modified.

Thirdly, the thermal energy density of the out-of plane movement of the graphene is larger than the van-der Waals energy density provided by the tip, if the tip atoms are more than 5 nm apart from the graphene (see eq.\ref{eq:vdW}). Thus, only a very small area of the graphene directly below the tip apex will be smoothed and stretched by $E_{\rm vdW,tip}$. We note in passing that previous MD simulations at $T=5$ K using a sphere for the representation of the tip found that only the inner 20 nm below the tip exhibit an inhomogeneous strain. This result was independent of the size of the sphere and the size of the membrane, if larger in radius than 20 nm or 200 nm, respectively. Thus, the exerted strain by the tip will appear on a rather reduced lateral size only.\\

Fourthly, we find experimentally that $I(z)$ decays much weaker on suspended areas than on supported areas (Fig. \ref{Fig1b}).
This contradicts the one-dimensional model displayed in Fig. \ref{Fig7}. There, an increased distance between tip and graphene results in reduced $E_{\rm vdW,tip}$, which implies a retraction of the membrane due to the unchanged retracting $E_{\rm strain}(h)$. This would inevitably lead to a stronger decay of $I(z)$ with vertical tip position $z$ than for a graphene layer fixed to the substrate, as observed, e.g., while lifting originally supported graphene.\cite{Mashoff}\\
 However, the van-der-Waals forces of the tip are strongly inhomogeneous in the tip area. One could imagine that the curvature of the Gaussian bump changes while retracting the tip. Consequently, the Gaussian gets smaller, and,thus, counterintuitively, the tunneling atom of the graphene could be moved towards the direction of the retracting tip. The low bending rigidity of graphene ($\kappa =1.4-1.6$ eV according to theory\cite{Wei} in agreement with experiment\cite{Tapaszto}) allows such a behavior in principle. We found a hint towards such a behavior by MD calculations of graphene on a SiO$_2$ substrate. The width of the pulled-up Gaussian indeed decreases from 7 \AA\hspace{1mm} to 6 \AA\hspace{1mm}, if the tip is moved away from the graphene by 0.5 \AA.\cite{Nemes} The Gaussian height does not get larger by retracting the tip, but it remains unchanged, which already contradicts the naive expectation that the Gaussian height decreases when retracting the tip. This surprising behavior was only found at tip-graphene distances of 3.5 \AA\hspace{1mm} for the graphene on the substrate, which is unreasonably small for $I=0.1-0.5$ nA. But a scenario with increased Gaussian height at retracting tip might appear at larger distance on suspended graphene due to the more flexible boundary conditions on the membrane. Eventually, extensive MD simulations using large suspended samples with adequate boundaries are required to confirm such a scenario.
 At the moment, we take it as the only scenario compatible with the experimental data.\\

 These four arguments led us to the model sketched in Fig. \ref{Fig6}.  The Gaussian bump with dynamically rippled, concave tails is then moved with the tip across the sample.

\subsection{Enlarged atomic resolution by drag}
Coming back to the measured apparent rippling, we should exclude that the interaction of the tip with the graphene membrane results in an extended lateral scale of the normal atomic corrugation. This is an appealing scenario since we do not observe atomic resolution at the correct scale on the membrane areas (Fig. \ref{Fig4}b).\\
 To test this scenario, we have performed numerical simulations using pairwise interaction potentials of Lennard-Jones type between graphene atoms and W atoms of a pyramidal tip with (110) axis. This allows to estimate the lateral corrugation of the strength of the van-der-Waals interaction between tip and graphene during scanning.\cite{Nemes} It firstly turns out  that $E_{\rm vdW, tip}\simeq 1$ eV at a typical tip-graphene distance of $\Delta  z=4-6$~\AA\hspace{1mm} in reasonable agreement with the estimates above.  More importantly, the lateral corrugation of $E_{\rm vdW, tip}$ is less than 1 meV, and, thus, well below $k_{\rm B}T$.\cite{Nemes} Consequently, the atomic lattice can not be dragged with the tip in tunneling distance at 300 K.\\
 In addition, we discuss, if an unlikely mechanical contact between tip and graphene can explain an enlarged image of atomic resolution. This resembles the situation of contact AFM, which exhibits apparent atomic resolution known to be caused by dragging the substrate atoms with the tip, until the lateral tension is large enough to overcome the favorable commensurate adaption between tip atoms and sample atoms.\cite{Hoelscher} However, the subsequent periodic relaxation and build up of the lateral stress always eventually leads to a force map reproducing the atomic lattice symmetry and length scale.\cite{Hoelscher} STM can probe the periodically changing force map in contact, too, if the tip touches the sample by an insulating layer. Then, the periodically changing force changes the tunneling barrier periodically.\cite{Klijn} However, even in such an unlikely case (see above), the periodicity would still reveal the atomic lattice period, but not an enlarged one. If the lateral scan size, which is required to achieve the critical lateral tension for periodic force imaging, is large, strong differences between different scan directions would appear\cite{Hoelscher} in contrast to the experiment (Fig.\ref{Fig2}b and c).\\
Thus, we do not find any reasonable scenario leading to a simple magnification of the atomic lattice by a factor up to 150.
\subsection{Frozen phonons by negative tension}
Xu et al. provided a different scenario for a possible large wavelength corrugation on suspended graphene via frozen flexural phonons.\cite{Xu2,Xu3}  It starts with a different explanation for the slow decay of $I(z)$ curves, which are observed similarly in their measurements on suspended graphene as in ours (Fig. \ref{Fig1b}).\cite{Xu2,Xu3} They argued that an increased current locally heats the graphene below the tip, which consequently contracts.\cite{NTA} Thus, the tip has to get closer to the graphene in order to maintain a tunneling current. However, the thermal conductivity of suspended graphene has been measured to be more than $\kappa _{\rm 3D}= 2000$~W/mK at room temperature and can be additionally increased by nuclear purification.\cite{Pop}
This corresponds to a 2D heat conductivity $\kappa _{\rm 2D}=\kappa _{\rm 3D}\cdot t= 0.6$~$\mu$W/K ($t=0.335$ nm: effective thickness of graphene). Assuming a circular geometry
of heat flow below the tip and a complete transformation of the maximum power of the tunneling electrons in our experiments ($I=0.5$ nA, $V=-0.3$ V) into heat, i.e., an exerted heat power of $P_{I}=1.5 \times 10^{-10}$ W, we obtain a maximum temperature increase of $\Delta T\simeq P_I/\kappa_{\rm 2D}\cdot r/2\pi r = 0.04$ mK independent on the temperature relaxation length $r$.
This is far below the temperature stability of the experiment and would lead to a negligible local contraction of the graphene regarding the thermal expansion coefficient at 300 K of $\alpha\simeq -7\times 10^{-6}$/K.\cite{NTA} Taking a typical aspect ratio of the bump below the tip of width/height = 10, we would require a contraction of about 30 nm  within an area of a few $100-1000$ nm below the tip in order to explain the $I(z)$ curve in Fig. \ref{Fig1b}. This would require local temperatures far above the melting temperature of graphene. Of course, $\kappa _{\rm 2D}$ could be decreased due to the stretching, which hardens the flexural phonons responsible for the large $\kappa_{\rm 2D}$, but surely not by the required more than eight orders of magnitude. Recall that the stretching appears only on a length scale of 5 nm. Moreover, the negative $\alpha$ is also based on the increased flexural movement with temperature\cite{NTA}, thus, being suppressed by stiffening the stretched membrane.
Consequently, we can safely exclude the local heating model for our measurements.\\
This is important, since an appealing explanation for a static rippling has been provided by partly the same authors.\cite{Xu3} By an analysis of a continuum model of an elastic membrane under heat load from the tip, they find a slow-down of flexural phonon frequencies towards a critical one. They use an elastic model which includes the contraction due to heating by the tunneling current. This reveals a negative surface tension favoring static ripples.
Since the negative tension increases with current, the critical wave length decreases with current being estimated to be 26, 18, and 8 nm at locally increased temperatures of 10 K, 20 K and 100 K, respectively. This would reproduce the trend of our experiments including the correct length scales. However, as pointed out above, it is extremely unlikely that such a huge temperature difference is induced by the tunneling current. Moreover, we believe that the model is inconsistent. A contraction of graphene by a temperature increase must stiffen the membrane and cannot lead to a weakening of the membrane eventually leading to frozen phonons.
\subsection{Standing phonon waves}
As a final scenario, we discuss the possibility of exciting a standing phonon wave. The tunneling current might stochastically excite flexural phonons by the force between the tunneling electron and its image charge. The excitation leads to oscillations of the membrane changing the tunneling probability periodically. If the period fits to the frequency $f$ of the tunneling electrons ($f=I/e$), the phonon frequency $f$ will be amplified such that the corresponding phonon will be finally selectively induced. Reflection of the phonon at the marker hole boundary might lead to a standing wave of that phonon such that the oscillation amplitude varies periodically in real space by $\lambda/2$ ($\lambda$: wave length of phonon). The non-linearity of $I(z)\propto e^{-\alpha\cdot z}$ naturally leads to a stronger DC current in areas of larger oscillation amplitude\cite{Mashoff} and, thus, to a slight retraction of the tip in the antinodal areas of the standing wave. The model is appealing since it reveals a decreasing wavelength with increasing current as found in experiment.\\
The largely isotropic dispersion of the large wavelength flexural phonons at 300 K $\omega_{\rm FP}(k)$ ($\omega_{\rm FP}$: frequency, $k$: wave vector) has been deduced from Monte-Carlo simulations in reasonable agreement with indirect experiments (heat conductivity).\cite{Ramirez} It consists of a largely
temperature-independent quadratic term and a strongly temperature-dependent linear term describing the anharmonic coupling to in-plane phonons:
\begin{equation}
\omega_{\rm FP}=\sqrt{\frac{\sigma}{\rho}k^2+\frac{\kappa}{\rho}k^4}
\end{equation}
Using room temperature values of bending rigidity $\kappa=1.6$ eV, density $\rho=7.6\cdot 10^{-7}$ kg/m$^2$ and stress parameter $\sigma=0.008$ eV/\AA$^2$,\cite{Ramirez}
 we  straightforwardly find the wave vector $k$ corresponding to $I=e\cdot\omega_{\rm FP}/2\pi$. They turn out to be $k=10^7$/m and $k=5\cdot10^7$/m at $I=0.1$ nA  and $I=0.5$ nA, respectively. This leads to $\lambda/2=\pi/k=300$~nm  and $\lambda/2=60$~nm, respectively, being about an order of magnitude too large (see Fig. \ref{Fig3}a). Even neglecting the linear term in the phonon dispersion reveals slightly too large values of $\lambda/2=\pi/k=40$ nm (17 nm). We expect, moreover, that the membrane is additionally stiffened and not weakened by the interactions with substrate and tip, which increases the resulting wavelengths further such that it gets even more incompatible with the experiment.\\

Thus, we have to conclude that there is currently no quantitatively correct scenario which can explain our data. The analysis is, of course, based on parameters, which are partly not completely settled experimentally, e.g., the low energy dispersion of the flexural phonons has not been measured directly. However, they might serve as an excellent starting point for further experiments and analysis. We hope that our preliminary analysis, also showing that temperature induced dynamical forces are decisive for the total energy balance, stimulates further experiments and theoretical modeling including large scale Monte-Carlo simulations eventually leading to a convincing explanation of this intriguing result. Experimentally, it would be desirable to use a controlled shape and chemistry of the tip crosschecked, e.g., by electron microscopy. Alternatively, one could stiffen the graphene membrane, e.g., by creation of vacancies\cite{Guinea}, by strongly reducing the distance between graphene and substrate, or by applying a backgate voltage.

\section{Summary}
In summary, by STM on suspended graphene more than 1 $\mu$ m away from the supporting edges, we have observed an apparent large scale corrugation with honeycomb symmetry and amplitude of about 100 pm changing in periodicity from 40 nm to 15 nm, if the tunneling current is increased from $I=0.1$ nA to $I=0.5$ nA. The appearance is independent of scan speed and scan direction and observed reproducibly after retracting the tip and reapproaching the membrane or after moving the tip to the supported areas, where usual STM images are recorded, and back onto the membrane. We argue that the observed pattern can neither be induced by an instrumental artifact nor by a measurement induced magnification of the atomic lattice, e.g., due to drag of the graphene by the tip. We also rule out that the local heating by the tunneling current is of importance. Finally, we discuss the induction of standing phonon waves by the tunneling current, which, however, appear to be larger in wavelength than found in the experiment. Without a convincing explanation, we hope that this intriguing result stimulates further STM work on and modeling of suspended graphene.\\

\section*{Acknowledgement}
We acknowledge helpful discussions with F. Guinea and F. Libisch as well as financial support by the
Graphene Flagship (Contract No. NECTICT-696656) and the German Science foundation (Li 1050-2/2 through SPP-1459).

%
%

\end{document}